# Evidence for Anion-Free-Electron Duality and Enhanced Superconducting Role of Interstitial Anionic Electrons in Electrides

Zhao Liu[1], Xiang Wang[1], Yin Yang[1], Pengcheng Ma[1], Zhijun Tu[2,3], Xinyu Wang[4], Donghan Jia[4], Wenju Zhou[4], Huiyang Gou[4], Hechang Lei[2,3*], Qiang Xu[5*], Zhonghao Liu[1*] and Tian Cui[1,6*]

[1]*Institute of High Pressure Physics, School of Physical Science and Technology, Ningbo University, Ningbo, 315211, People's Republic of China*

[2]*School of Physics and Beijing Key Laboratory of Opto-electronic Functional Materials & Micro-nano Devices, Renmin University of China, Beijing 100872, China*

[3]*Key Laboratory of Quantum State Construction and Manipulation (Ministry of Education), Renmin University of China, Beijing, 100872, China*

[4]*Center for High Pressure Science and Technology Advanced Research, Beijing, 100193, China*

[5]*Key Laboratory of Material Simulation Methods & Software of Ministry of Education, College of Physics, Jilin University, Changchun, 130012, China*

[6]*State Key Laboratory of High Pressure and Superhard Materials, College of Physics, Jilin University, Changchun, 130012, People's Republic of China*

**ABSTRACT**

The discovery of superconducting electrides, characterized by interstitial anionic electrons (IAEs) residing in lattice cavities, has established a distinctive platform for investigating superconductors. Yet the superconducting origin and the fundamental role of IAEs in Cooper pairing formation remain poorly understood due to the challenges in directly observing IAEs. Here, combining angle-resolved photoemission spectroscopy (ARPES), transport measurements, and first-principles calculations, we certify that the IAEs in electride La$_3$In ($T_c$ = 9.4 K) exhibit a dual nature as both anions and free electrons. With the finite-depth potential well model, we trace that IAEs originate from electronic states near the Fermi level located above potential barriers, forming a Fermi sea susceptible to scattering by La-derived phonons, triggering superconductivity. ARPES combined with high-resolution XRD measurements on oxygen-treated samples directly reveals IAEs' spatial distribution and energy dispersion from interstitial sites with the consistent energy value predicted by our theory model. The concomitant diminution of free electrons upon oxygen treatment, leading to a marked reduction in superconductivity, further provides compelling experimental evidence that IAEs actively participate in electron-phonon coupling. Our findings resolve the long-standing ambiguity regarding the electronic nature of IAEs, elucidate their enhancing superconductivity in the phonon-mediated mechanism, and provide a foundation for exploring advanced electride-based superconductors.

Corresponding Authors

Email: hlei@ruc.edu.cn; xuq@jlu.edu.cn; liuzhonghao@nbu.edu.cn; cuitian@nbu.edu.cn

**Introduction**

The quest for new types of superconductors and the elucidation of their superconducting mechanisms remains one of the most important research subjects in physics and materials science [1-3]. Superconducting electrides, materials in which interstitial spaces host interstitial anionic electrons (IAEs), have recently been recognized as a new class of conventional BCS-Eliashberg superconductors [4-6]. Among the early milestone developments are the experimental realization of electride $(Ca_{24}Al_{28}O_6)_4^+(e^-)_4$ exhibiting an observed superconducting transition temperature ($T_c$) of 0.2 K, along with $Mn_5Si_3$-type electrides $Zr_5Sb_3$ and $Nb_5Ir_3$ displaying $T_c$'s in the range of approximately 2.3 to 9.4 K at ambient conditions [7]. Currently, a series of high-pressure electrides have been predicted to possess critical temperatures exceeding the McMillan limit or even surpassing the temperature of liquid nitrogen, such as $Li_6P$ with 41.36 K at 200 GPa [8], $Li_8Au$ with 73 K at 250 GPa [9], and $Li_4Rh$ with 108.2 K at 300 GPa [10]. However, the fundamental role of IAEs in facilitating or enhancing superconductivity remains deeply controversial and constitutes a significant knowledge gap. Indeed, the classic view believes that the IAEs in electrides $Li_3S$ [11], $Li_5Si$, $Li_5Sn$ [12], and $Li_6C$ [13] consistently exhibit a negligible or even suppressing role in superconductivity, which contrasts sharply with the promoting effect proposed in $Li_6P$ [8,14]. This discrepancy highlights the ambiguity surrounding the superconducting mechanism in electrides, which is further compounded by the formidable experimental challenges in directly probing the spatial behavior and energetic distribution of IAEs. Consequently, a unified physical picture characterizing the existence state of IAEs and explaining the interplay between IAEs and superconductivity is lacking.

The key to resolving this impasse lies in understanding the electronic nature of IAEs, especially in relation to the mechanism of Cooper pair formation. According to the BCS theory, the formation of the Fermi sea characterized by Cooper pairs is regarded as a necessary condition for achieving the superconducting transition [15]. Verifying whether IAEs exhibit free-electron behavior is a crucial precondition for unraveling the superconducting causation. Given the generation progress of IAEs, there are currently two prevailing perspectives: Dong and Organov attempted to account for the Gaussian-like density distribution of IAEs by applying the uncertainty principle within a confined interstitial volume [16]. Another model suggests that the high pressure elevates the energy levels of atomic orbitals above those of IAE-quantized orbitals, thereby initiating charge transfer and resulting in the formation of localized regions with high charge density [17]. All these viewpoints share a common physical picture: confined interstitial potential wells trigger the formation of bound-state IAEs. Conversely, recent theoretical studies increasingly suggest that IAEs behave as free electrons near the Fermi level ($E_F$) [18,19]. Consequently, achieving consistency between theoretical insights and experimental observations in revealing the IAEs states and their connection to superconducting pairing is now at the forefront of research.

In this work, we bridge this gap by combining angle-resolved photoemission spectroscopy (ARPES), transport measurements, and first-principles calculations on high-quality single crystals of the electride La$_3$In. Our transport measurement confirmed that La$_3$In exhibits superconductivity with a $T_c$ of approximately 9.4 K, along with T-linear resistivity over a wide temperature range. We propose the finite-depth potential wells model directly elucidating the anion-electron duality of IAEs and provide evidence for IAEs acting as free-electron behaviors. Further electron-phonon coupling (EPC) calculations reveal that such a high-$T_c$ is closely associated with the IAEs being scattered by La-derived phonons. Importantly, single-crystal X-ray diffraction demonstrates that oxygen atoms, introduced via oxygen treatment, occupy the body-centered sites – precisely in the interstitial anionic positions – while ARPES directly resolves the corresponding oxygen-derived bands, whose energy positions are in excellent agreement with the predictions of our finite-depth potential well model. Additionally, the diminution of free electrons upon oxygen treatment leads to a marked reduction in superconductivity, which offers a direct experimental confirmation that IAEs actively participate in EPC. Our study confirms the causal relationship between IAEs and high-temperature superconductivity, providing a reliable basis for the exploration of high-$T_c$ superconductors in electrides.

## Results

### Phonon-mediated superconductivity

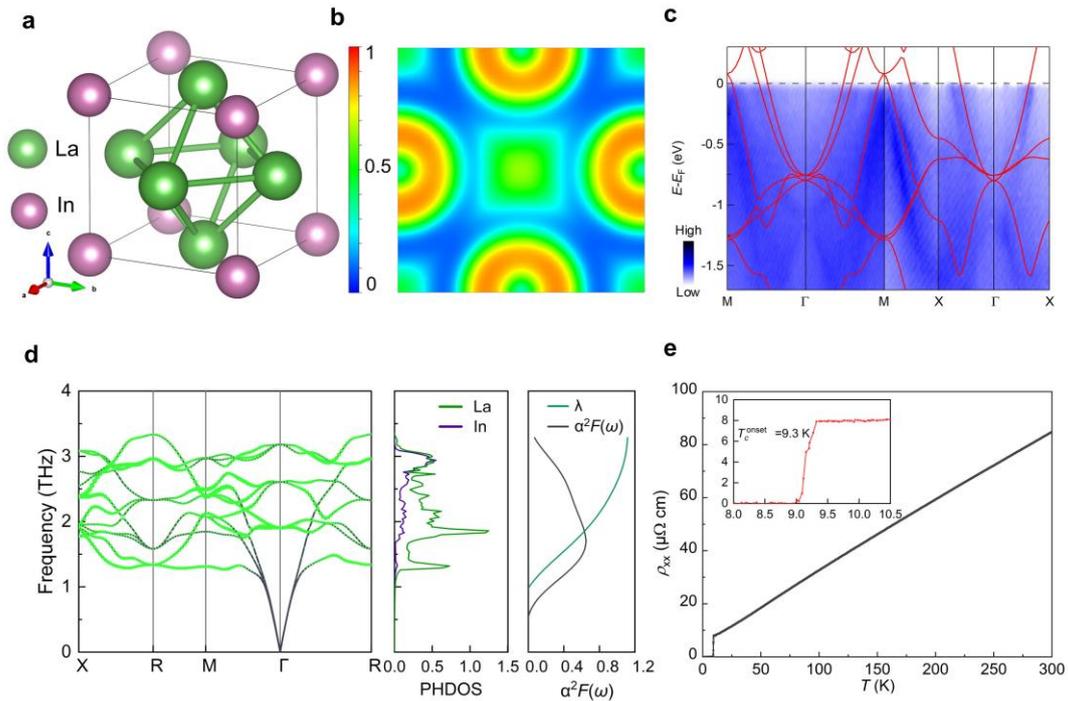

Fig. 1. (a) The *Pm*-3*m* phase of La$_3$In at ambient conditions. (b) 2D isosurface of electron localization function (ELF). (c) ARPES intensity plots along high-symmetry lines appended with the calculated bands without any renormalizations and shifts. (d) Phonon dispersion, projected phonon DOS, and $\alpha^2F(\omega)$ with integral $\lambda(\omega)$ at ambient conditions. The solid blue dot on the phonon spectra represents the contribution of the phonon linewidth in proportion to the spherical

scale. (e) Temperature dependence of $\rho_{xx}$(T) at zero magnetic field. Inset picture: zooming in on the translation temperature region.

We conducted the La-flux method to grow La$_3$In single crystals and confirmed its *Pm*-3*m* structure by single-crystal X-ray diffraction (XRD) [20]. The Bravais lattice, depicted in Figure 1(a), contains one formula unit of La$_3$In, with La and In atoms occupying the vertices 3*c* (–0.500 –0.500 –0.000) and center 1*a* (–0.000 –0.000 –0.000) Wyckoff positions, respectively (see the supplementary table S1). The cavity volume within an octahedral cone formed by La atoms with an equal atomic distance of 3.592 Å is close to 21.848 Å$^3$. Our ELF analysis in Figure 1(b) revealed that the IAEs are primarily localized at the lattice center. Further coordination environment analysis demonstrated that this structure comprises six-coordinated IAEs within the La cation. Figure 1(c) shows the low-energy dispersions of La$_3$In measured by ARPES along the high-symmetry directions, overlaid with the calculated bands without any renormalization or energy shifts. The experimental data agree well with the calculations, except for the bands near –1 eV below the $E_F$, which require a downward shift. This discrepancy can be attributed to interband coupling and correlation effects [21,22]. A prominent feature is the large electron-like band surrounding the Γ point, shaping a parabolic function characteristic, confirming the electron-type nature of the carriers and reflecting the behavior of free electrons based on Thomas-Fermi theory [23]. Moreover, a Van Hove singularity at the M point markedly enhances the density of states (DOS) at the $E_F$; together with correlation effects, this may promote superconductivity and lead to strange-metal behavior as reflected in the T-linear resistivity shown in Figure 1(e) [20].

Next, we calculate the Allen–Dynes–modified McMillan equation to demystify the superconducting origin of *Pm*-3*m* phase. The $T_c$, projected phonon DOS, and $α^2F(ω)$ with integral $λ(ω)$ of the *Pm*-3*m* phase are displayed in Figure 1(d). An estimated $T_c$ of 9.5 K under ambient conditions was validated to be consistent with the experimental result [see Fig. 1(e)]. Based on the BCS theory, the essential factors for phonon-mediated superconductivity are especially expounded, including: (i) free electrons near $E_F$, (ii) characteristic phonon frequency, and (iii) the strength of EPC. On the aspect of phonon DOS in Figure 1(d), the La-associated acoustic phonon vibrations strongly couple with electrons, contributing 72% to the total EPC. Remarkably, these acoustic branch phonons with larger frequencies afford a broader range of energy from various vibrational modes, enhancing the likelihood of satisfying the formation energy of Cooper pairs [19,24], which are reflected by the large projected phonon linewidths, integral EPC strength $λ(ω)$ of 2.72, and a high peak in the Eliashberg $α^2F(ω)$.

**Generation mechanism and existence state of IAEs**

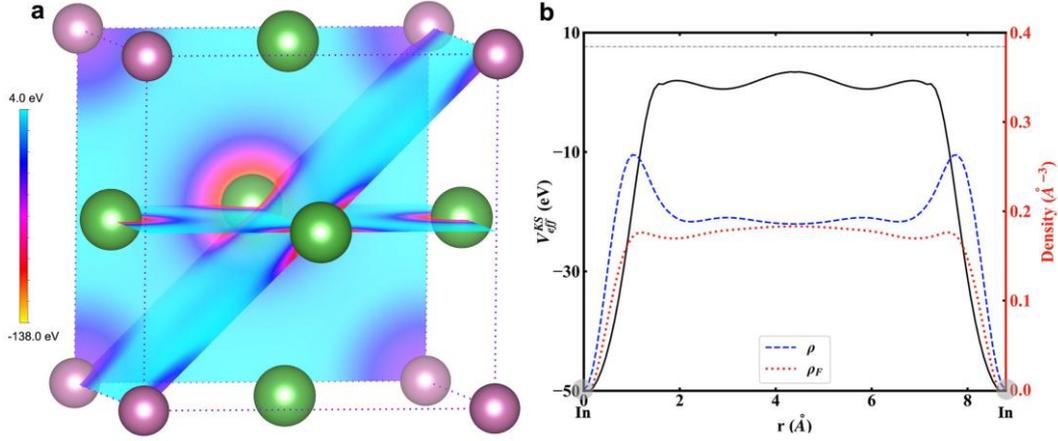

Fig. 2. (a) The distribution of Kohn-Sham effective potential, including contributions from the ionic, Hartree, and exchange-correlation potentials, in the *Pm*-3*m*-La$_3$In system. It depicts the energy values associated with the occurrence of effective local potential and the maximum of the local potential barrier at the center and body-centered positions of the lattice. (b) The effective local potential $V_{eff}^{KS}$, total electron density ($\rho$), and free-electron density ($\rho_F$) along [111] direction of La$_3$In.

To elucidate the physical picture underlying the existence state of IAEs, we conducted the effective local potential simulations within the framework of Kohn-Sham density functional theory. Intriguingly, Figure 2(a) confirmed that the IAEs are localized at the maximum position of the potential within the interstitial region. This novel phenomenon starkly contrasts with the conventional expectation that IAEs are confined to interstitial potential wells [16,17,25], but instead exemplifies a typical counterintuitive "potential-barrier affinity effect", recently been proposed and elucidated in our study [26]. This "potential-barrier affinity effect" is manifested as the interstitial electron accumulation phenomenon when their energy exceeds the potential-barrier height. These results have been further corroborated in Figure 2(b), which displays the Kohn-Sham effective local potential $V_{eff}^{KS}$, total electron density $\rho$, and the free electron density ($\rho_F$) corresponding to unbound-states along [111] direction of La$_3$In. Observations confirm that the electron density around the peak potential is overwhelmingly dominated by the $\rho_F$, whereas the electron density component ($\rho$−$\rho_F$) confined in bound-states is negligible. Such an abnormal "potential-barrier affinity effect" promotes the formation of IAEs in the interstitial barrier region of electrides, leading to a large number of electrons accumulated at the maximum point of the potential barrier, naturally forming anions. Meanwhile, this result clearly reveals that the IAEs refer to the electrons distributed within the energy range from the maximum effective potential $E_M$ to $E_F$.

**Intrinsic characteristics of IAEs performing free-electron behavior as a Fermi sea**

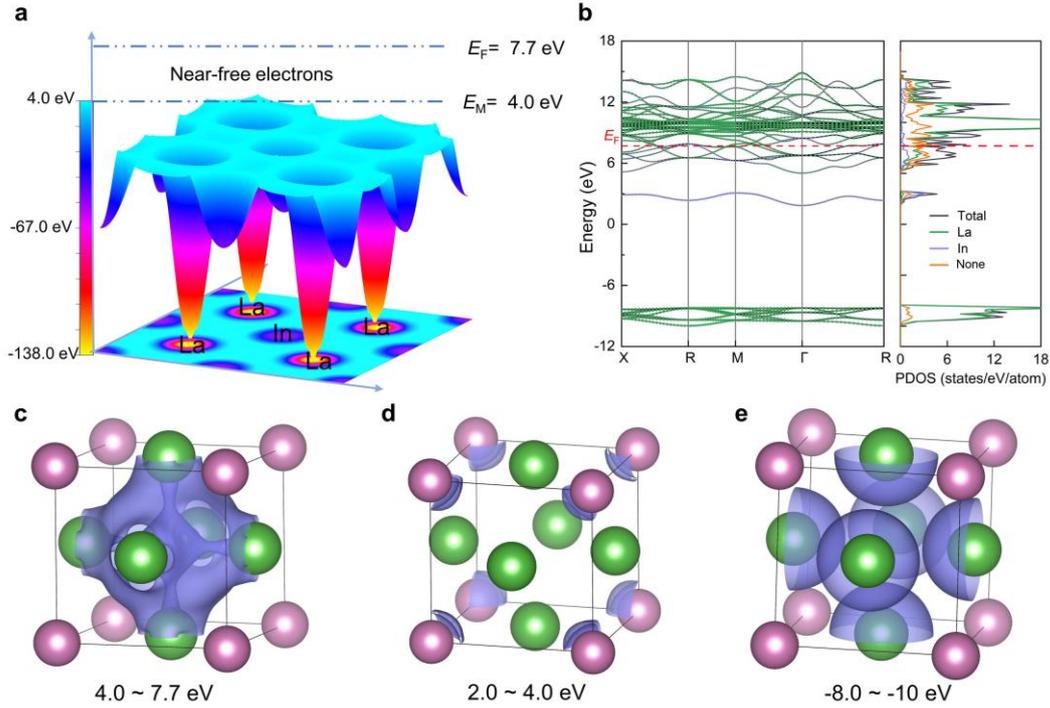

Fig. 3. (a) The section of 3D and 2D potential surface of Kohn-Sham effective local potential of La$_3$In. (b) The orbital-resolved electronic band structure of $Pm$-$3m$-La$_3$In phase. (c-d) The partial density of states (PDOS), where the partial charge density maps with an isosurface of 0.012 correspond to energy bands within different energy ranges.

To verify whether IAEs exhibit free-electron behaviors, the landscape of effective local potential around selected La atoms and the corresponding calculated band structures are shown in Figure 3(a–b). Notably, the $E_F$ = 7.7 eV lies significantly above the maximum effective potential ($E_M$ = 4.0 eV). These results suggest a pronounced near-free electron behavior across a broad range of occupied states, spanning from 4.0 eV to 7.7 eV, which provides the necessary conditions for the Fermi sea required in superconductivity. The formation mechanism of the free electrons is attributed to strong Pauli repulsion among electrons results in the occupation of higher energy levels. As a result, the higher kinetic energy of electrons here exceeds the limit of effective potential wells formed by cations of La$^{+2.32}$ and In$^{+0.32}$ (see the supplementary table S2), causing partial electrons to be expelled from these potential wells, performing near free-electron behaviors. In fact, the behavior of free electrons reflected within the energy range from the maximum effective potential $E_M$ to $E_F$ represents another state of existence of IAEs in the interstitial region–the free-electron behavior–as shown in the electron density contributed by the $E_M$ to $E_F$ near free electron band in Figure 3(b). Based on the ionic explanation of the interstitial electrons accumulation phenomenon presented in the previous text, we elucidate that IAEs indeed exhibit an anion-free-electron duality.

Subsequently, we calculate the corresponding projection of partial charge densities onto the constituent energy band to further confirm the free-electron behaviors of IAEs, thereby determining whether these states originate from bound or unbound states. Within the free-electron energy range of 4.0 eV to 7.7 eV characteristic of unbound states, a distinct charge distribution dominated by IAEs constitutes the corresponding energy bands as shown in Figure 3(c), confirming that the IAEs exhibit a clear free-electron feature. In the bound-state bands from 2.0 to 4.0 eV, $< E_M$, it can be observed that the charge distribution occupying the energy band mainly consists of components of In-$p$ orbitals at the vertex. At deeper energy levels of −9.0 eV to −8.0 eV, these energy bands are mainly composed of La-$d$ orbitals without IAEs occupied. However, the absence of IAEs at the center of the crystal cavity further supports the conclusion that IAE is not confined to the potential well. These results of energy band analysis are consistent with our proposed effective potential well model, confirming that the IAEs are not confined to traditional potential well models to exhibit a high charge density within the crystal cavity. Furthermore, the direct evidence for the prominent role of IAEs behaving as near-free electrons clearly reflects the crucial role of IAEs in contributing to the formation of free electrons as a Fermi sea.

In fact, as illustrated in Figure 3(b, c), the local orbital analysis is insufficient to capture the behavior of free electrons, as it represents only a partial component of the complete free-electron wave function. Above the maximum effective potential (> 4 eV), the charge distribution *None* (*None* refers to the components of the wave function that are not projected onto the atomic orbital) and La-$d$ electrons all belong to the interstitial free electrons IAEs because the projected DOS solely reflects the quantitative contribution of these interstitial electrons, projected onto various atomic orbitals. Notably, the electronic states above $E_M$ form strongly phonon-coupled bands as reflected by bands hopping of approximately 3.7 eV in Figure 3(a, b). These near-free electronic states have a strong coupling with the fundamental vibrations of constituent La atoms, as evidenced by the projected EPC strength in phonon spectra. Instructively, current findings of La$_3$In with IAEs and high-$T_c$ reflect the strong correlation between IAEs and superconductivity.

**Evidence of IAEs distribution above potential barriers and enhanced superconductivity**

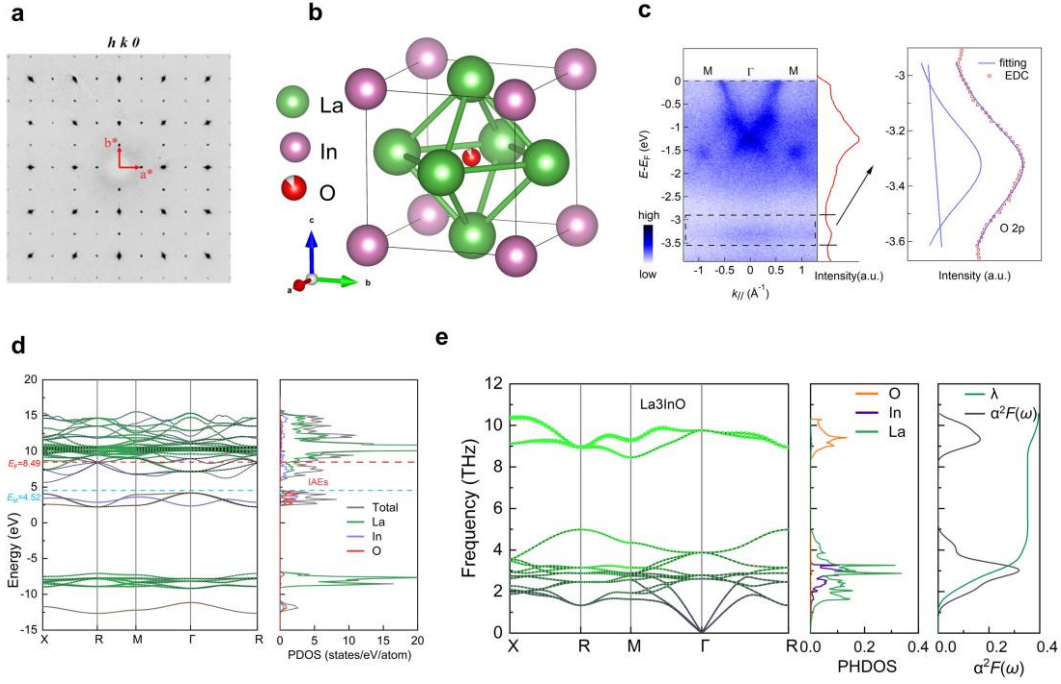

Fig. 4. (a) The reconstructed reciprocal lattice plane of $La_3InO_{0.9}$ with (b) its corresponding crystal structure resolved by high-resolution XRD. (c) ARPES intensity plot along the Γ–M direction shows O $2p$ orbital near -3.33 eV below $E_F$ for $La_3InO_{0.9}$. The energy distribution curve at the Γ point is shown in the right panel. The peak is fitted by the Lorentz function with a linear background. (d) Electronic band structure and projected DOS for $La_3InO$. (e) Phonon band structure information and corresponding $\alpha^2F(\omega)$ with integral $\lambda(\omega)$ for $La_3InO$.

To obtain experimental insight into the existence state and formation mechanism of IAEs, which cannot be directly resolved by ARPES, we performed oxygen treatment on $La_3In$ samples. Because oxygen atoms naturally occupy the interstitial sites where IAEs form, this approach allows us to identify the IAE positions in real space and to detect the corresponding oxygen-derived electronic states in momentum space. As shown below, the oxygen-treated samples reveal that IAEs originate from non-bound electronic states rather than from conventional bound-state confinement, and they further confirm the essential role of IAEs in enhancing superconductivity.

Herein, the reconstructed reciprocal lattice plane shown in Fig. 4(a), obtained from single-crystal XRD analysis, reveals the crystal structure of oxygen-treated $La_3InO_{0.9}$ and clearly demonstrates that oxygen atoms occupy the body-centered sites, precisely corresponding to the interstitial anionic positions (See the supplementary table 4). ARPES measurements directly resolve the oxygen-derived bands at approximately -3.3 eV below the $E_F$ [see Figure 4(b)], with energy positions in good agreement with the predictions of our finite-depth potential well model (See the supplementary Figure S1). Through applying oxygen treatment to the sample of $La_3In$ and examining the oxygen occupancy sites using single-crystal XRD, we present concrete evidence for the existence of IAEs in crystal cavities. Because it is hard to experimentally control the oxygen contents in samples, we further simulated various oxygen content doping scenarios ($La_3InO_x$), as shown in Figure S1, and consistently observed that with increasing oxygen doping, the free electron density from the maximum effective potential $E_M$ to the $E_F$ decreased, thereby confirming that the IAE energy states are indeed located above the potential barrier.

To further elucidate the role of IAEs to superconductivity, in particular, we simulated the oxygen content percentage occupying IAE sites in Figure 4(d, e), where the oxygen-induced bound states in La$_3$InO restrict the availability of approximately two free electrons. This is confirmed by integrating the electron count from $E_M$ = 4.51 eV to $E_F$ = 8.49 eV, as illustrated in Figure S2. These formal two free electrons, originally distributed within the $E_M$ – $E_F$ energy range, become confined due to the potential wells created by oxygen atoms. Consequently, the $T_c$ of La$_3$InO decreases to only 0.5 K. Additionally, combined calculations of the energy band structure and phonon spectrum reveal that oxygen atoms do not form covalent bonds with neighboring La atoms, indicating their negligible contribution to the phonon-mediated EPC. This result suggests that the IAE-associated free electron behavior plays a crucial role in Cooper pair formation through EPC.

**Discussion**

In summary, we bridge the experimental evidence gap of superconducting origin, existence state and the fundamental role of IAEs in Cooper pairing formation through integrated ARPES, transport measurements, and first-principles calculations on high-quality single crystals of the electride La$_3$In. Transport measurements confirm superconductivity in La$_3$In with a transition temperature $T_c \approx$ 9.4 K, accompanied by $T$-linear resistivity across a broad temperature range. We propose a finite-depth potential well model that directly captures the anionic-electron duality of IAEs and provides evidence for their free-electron behavior. EPC calculations further reveal that the elevated $T_c$ arises from scattering of IAEs by La-derived phonons. Importantly, single-crystal X-ray diffraction shows that oxygen atoms − incorporated via oxygen treatment − occupy body-centered sites, which correspond precisely to the interstitial anionic positions. ARPES data directly resolve oxygen-derived bands at energies consistent with our model predictions. Moreover, oxygen treatment reduces the free-electron concentration and sharply suppresses superconductivity, offering direct experimental evidence that IAEs participate actively in EPC. Our study not only but also reveals the promoting effect of IAEs on EPC, which has extensive implications for comprehending the origins of superconductivity and exploring high-$T_c$ superconducting electrides.

## Conflicts of interest

The authors declare no conflict of interest.

## Author Information

Corresponding Authors

Email: hlei@ruc.edu.cn; xuq@jlu.edu.cn; liuzhonghao@nbu.edu.cn; cuitian@nbu.edu.cn

# Acknowledgments

This work was supported by the National Key Research and Development Program of China (Grants No. 2023YFA1406200, No. 2022YFA1405500, No. 2023YFA1406500, No. 2022YFA1403800), National Natural Science Foundation of China (Grants No. 12304021, No. 52072188, No. 12222413, No. 12274459, No. 12305002, No. 12074013), National Science Fund for Distinguished Young Scholars (Grant No. T2225027), Zhejiang Provincial Natural Science Foundation of China (Grant No. LQ23A040004), Natural Science Foundation of Ningbo (Grant No. 2022J091, No. 2024J019), the Natural Science Foundation of Shanghai (Grants No. 23ZR1482200), the funding of Ningbo Yongjiang Talent Program and the Mechanics Interdisciplinary Fund for Outstanding Young Scholars of Ningbo University (Grants No. LJ2024003), Program for Science and Technology Innovation Team in Zhejiang (Grant No. 2021R01004), Program for Changjiang Scholars and Innovative Research Team in University (No. IRT_15R23). The calculations were performed in the Supercomputer Center of NBU.

**Single-crystal synthesis, characterizations, ARPES measurements and calculation methods**

High-quality single crystals of $La_3In$ were synthesized by the self-flux method [20]. Electrical transport measurements were conducted using a superconducting magnetic system (Cryomagnetics, C-Mag Vari-9). The crystal was then mounted on a goniometer and measured on a Rigaku XtaLAB Synergy R four-circle diffractometer with Mo Kα radiation (λ = 0.71073 Å) at room temperature. To ensure high data completeness, a multi-scan data collection strategy was employed using ω-scans with a frame width of 0.5° and an exposure time of 3s per frame. Data processing, including reduction and absorption correction, was performed using CrysAlis$^{Pro}$ (Rigaku Oxford Diffraction, 2024). The structure was solved using Olex2 (Dolomanov et al., 2009)[27] and refined by full-matrix least-squares on $F^2$ with SHELXL (Sheldrick, 2015)[28]. All structural visualizations were generated by VESTA (Momma & Izumi, 2011)[29].

ARPES measurements were conducted at the 03U and Dreamline beamlines of the Shanghai Synchrotron Radiation Facility (SSRF). The energy and angular resolutions were better than 15 meV and 0.2°, respectively. Samples smaller than 1×1 mm$^2$ were cleaved *in situ*, yielding flat mirror-like (001) surfaces. During the measurements, the temperature was maintained at approximately 10 K, and the pressure was kept below $6.5×10^{-11}$ Torr.

The first-principles calculations were performed on VASP code using the projector augmented wave (PAW) method and Perdew-Burke-Ernzerhof (PBE) functional within the Kohn-Sham density functional theory[30,31]. The kinetic cutoff energy of 800 eV and Monkhorst-Pack ***k***-meshes with a grid spacing of 0.10 Å$^{-1}$ were then employed to ensure the self-consistent field tolerance of $0.1 × 10^{-4}$ eV/atom. The lattice-dynamical and superconducting properties were estimated using density functional perturbation theory (DFPT), which was implemented in the Quantum-

ESPRESSO code[32,33]. Ultrasoft pseudopotentials were used with a kinetic energy cutoff of 100 Ry, and the charge density was integrated on a centered 18 × 18 × 18 **k**-point mesh. The first-order potential perturbation and dynamical matrices were calculated using DFPT on an irreducible 6 × 6 × 6 centered **q**-point mesh. The Allen-Dynes modified McMillan equation is:

$$T_c = \frac{\omega_{\log}}{1.2}\exp\left[-\frac{1.04(1+\lambda)}{\lambda-\mu^*(1+0.62\lambda)}\right],$$

where $\lambda$ is the EPC parameter, $\omega_{\log}$ is the logarithmic average frequency and $\mu^*$ is the Coulomb pseudopotential parameter.

$$\alpha^2 F(\omega) = \frac{1}{2\pi N(E_f)}\sum_{qv}\frac{\gamma_{qv}}{\omega_{qv}}\delta(\omega-\omega_{q,v})$$

$$\gamma_{qv} = \pi\omega_{qv}\sum_{mn}\sum_{k}|g_{mn}^v(\boldsymbol{k},\boldsymbol{q})|^2\delta(\varepsilon_{m,k+q}-\varepsilon_f)\times\delta(\varepsilon_{n,k}-\varepsilon_f)$$

$$\lambda = 2\int d\omega\frac{\alpha^2 F(\omega)}{\omega}\;;\quad \omega_{\log} = \exp[\frac{2}{\lambda}\int\frac{d\omega}{\omega}\alpha^2 F(\omega)\ln(\omega)]$$